\documentclass[12pt,a4paper]{revtex4}
\usepackage{epsfig}
\begin{document}
\title{Motional dispersions and ratchet effect in inertial systems}
\author{W.~L.~Reenbohn$^{1}$, S.~Saikia$^{1,2}$, R.~Roy$^{3}$, Mangal ~C.~Mahato$^{1}$} 

\address{$^{1}${Department of Physics, North-Eastern Hill University, 
Shillong 793003, India}}
\address{$^{2}${St.Anthony's College, Shillong-793003, India}}
\address{$^{3}${Women's College, Shillong-793003, India}}

\begin{abstract}Abstract:~~We obtain ratchet effect in inertial structureless systems in 
symmetric periodic potentials where the asymmetry comes from the nonuniform 
friction offered by the medium and driven by symmetric periodic forces. In 
the adiabatic limit the calculations are done by extending the matrix 
continued fraction method and also by numerically solving the appropriate 
Langevin equation. For finite frequency field drive the ratchet effect is 
obtained only numerically. In the transient time scales the system  shows 
dispersionless behaviour as reported earlier when a constant force is applied. 
In the periodic drive case the dispersion behaviour is more complex. In this 
brief communication we report some of the results of our work.
\end{abstract}

\maketitle
\maketitle

\noindent
\newcommand{\nwc}{\newcommand}
\nwc{\bdm}{\begin{displaymath}}
\nwc{\edm}{\end{displaymath}}
\section{Introduction}
The phenomenon of obtaining net unidirectional current in a periodic potential 
without the application of any time averaged external field is 
termed as ratchet effect\cite{Reim}. This is necessarily a non-equilibrium 
phenomenon 
and has been investigated extensively in systems where damping to periodic 
motion is large. Also, in majority of the investigations the potential is 
considered asymmetric (and hence the name ratchet). In order to obtain ratchet 
effect in asymmteric potentials the system is rocked\cite{Magn} periodically 
(rocked ratchet) or the potential amplitude is changed\cite{Pros} 
dichotomously, periodically or randomly, (flashing or fluctuating-potential 
ratchets). The effect can also be obtained if the system is driven 
periodically but time asymmetrically in such a way that the total applied 
force per period is zero\cite{Mah1}. In such systems the condition for 
asymmetric periodic potential can be waived. All these popular models 
consider the systems to be homogeneous where the friction coefficient is 
taken to be constant in space and the temperature is maintained uniform. In 
the present work we consider an underdamped system and the periodic potential 
is taken to be symmetric. However, the friction coefficient is considered 
to vary 
periodically in space similar to the potential but with a phase difference. 
Also, the system is driven periodically symmetrically about zero. The 
overdamped case of the problem has been studied earlier\cite{But1}. 

Since the system asymmetry is solely due to non-uniformity of friction 
coefficient the expected net particle current in such a system may be weak
\cite{Mah2}. Nevertheless, the physical explanation for asymmetric particle 
current 
can be easily understood. In fig.1 the periodic potential $V(x)=-sin(x)$ and 
the friction coefficient $\gamma(x)=\gamma_0[1-\lambda sin(x+\phi]$ are 
plotted together with $\lambda = 0.9$, phase difference $\phi = 0.35$ and 
$\gamma_0=1$. In the figure a period of the potential is shown divided into 
two halves one with friction coefficient small ($<1.0$) and the other with 
large ($>1.0$) friction coefficient. The effect of $\phi$ on this division is 
clearly visible: The friction is more on the left of the potential peak 
position than on the right. 

The physical explanation for the possibility of obtaining ratchet effect 
in this inhomogeneous system can be given in two ways. Firstly, since in the 
static situation the position 
probability distribution is independent of the profile of the friction 
coefficient, the distribution will be Gaussian-like and symmetric about the 
minima of the potential. However, the situation changes in the dynamic case. 
Because of the higher friction on the left of the potential peak position the 
particle spends more time there during its motion. And hence the effective 
position probability distribution is skewed with higher probability on the 
left than on the right of the potential peaks as though a static constant force 
has been applied\cite{But1} in the left (negative) direction. Thus, one would 
expect a negative average particle current even without the application of 
net external field. Secondly, since the
\begin{figure}[t]
\epsfxsize=9cm
\centerline{\epsfbox{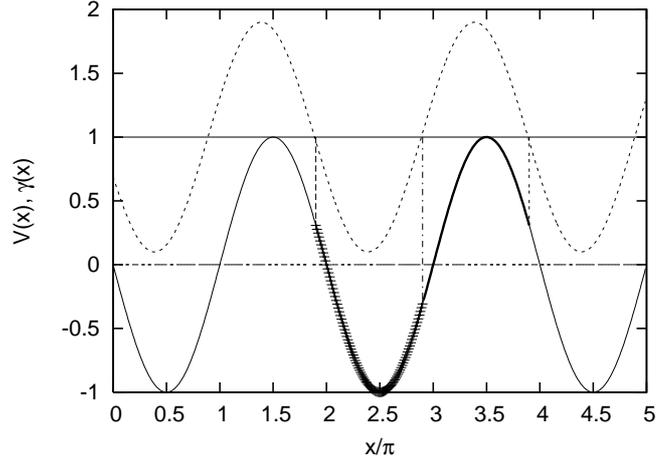}}
\caption{Plot of potential $V(x)$ (solid line) and the friction coefficient
$\gamma (x)$ (dashed line). The portion of a period of $V(x)$ where
$\gamma (x)<1$ is shown hatched and the portion where $\gamma (x)>1$ is 
shown by a thick line.}
\label{fig.1}
\end{figure}
particles tend to spend more time on the left of the peak it will absorb more 
thermal energy from the bath and hence effectively it will be at a higher 
temperature on the left side of the potential peaks. Higher effective 
temperature on the left leads to higher probability of thermally assisted 
passage from the left of the potential peak to the right giving rise to a net 
current in the right (or positive) direction. These two opposing effects in the
nonequilibrium situation may lead to competition in deciding the direction of 
the ratchet current depending on the temperature. These two competing effects 
could even lead to ratchet current direction reversals as will be revealed by our numerical results.

The underdamped systems are difficult to study (analytically as well as 
numerically) but provide interesting results. If the particle is initially put 
at the bottom of one of the wells of the potential (not with great initial 
velocity) it will get out of the well due to thermal effect (Kramers-like 
behaviour). However, once out of the well it will roll down from the top to 
the bottom of the next well of the tilted periodic potential without much 
frictional loss of energy and hence may continue to move along for a long 
distance (more than one period) before it again gets trapped in a distant well.
From there it will again get out due to thermal effects and the process will 
keep repeating. The distribution of travel distances between two consecutive 
haltings is itself quite interesting to study. When taken an ensemble of such 
particles they may collectively show coherence properties in the intermediate 
time scales, where position dispersions $<(\Delta x(t))^2>=<x(t)^2>-<x(t)>^2$
become constant in time for a brief period. This is how the long time steady 
state behaviour $<(\Delta x(t))^2 >\sim t$ is bridged with the very short time 
behaviour $<(\Delta x(t))^2>\sim t^{2}$. This result has been reported 
recently for 
a constant applied force\cite{Linde}. In the periodically driven case, however, 
the system shows more complex behaviour to be described as we proceed.

Obtaining analytical solutions of the Langevin equation for underdamped 
inhomogeneous (space-dependent friction) systems subjected to constant 
uniform force field in the steady state is in principle a straightforward 
extension of the matrix continued fraction method (MCFM) initially developed 
by Risken\cite{risk}. Our extended formalism, however, yields stable solutions 
only in a limited range of the applied force, especially for small friction 
coefficients. It turns out that the range of the applied force where MCFM 
yields sensible results is the one relevant to our discussion. In section II, 
we simply describe how to calculate the mean particle velocity in the steady 
state. However, it has not yet been possible to write down a similar 
expression for the solution of the Langevin equation when the system is 
subjected to a time varying external field. We, therefore, resort to numerical 
methods. The efficacy of the numerical method is established by comparing the 
results so obtained with the MCFM result in the adiabatic case (Fig.2). The 
numerical results along with the MCFM (adiabatic case) results are presented 
in section III. In section IV we discuss the salient features of our work to 
conclude.

\section{The Method }

Consider a particle of mass $m$ moving in a periodic potential 
$V(x)=-V_0 sin(kx)$ in a medium where the particle experiences friction 
with coefficient $\gamma (x)=\gamma_0(1-\lambda sin(kx+\phi))$ with $\lambda 
<1$. The motion
can be described by the following Langevin equation:
\begin{equation}
m\frac{d^{2}x}{dt^{2}}=-\gamma (x)\frac{dx}{dt}-\frac{\partial{V(x)}}{\partial
x}+F(t)+\sqrt{\gamma(x)T}\xi(t),
\end{equation}
where $T$ is the temperature in energy units in terms of $k_B$, the Boltzmann
constant and $F(t)$ is the externally applied field. $\xi (t)$ is the 
fluctuating force term with $<\xi (t)>=0$, 
and $<\xi (t)\xi(t^{'})>=2\delta(t-t^{'})$.
In dimensionless units with $m=1$, $V_0=1$, $k=1$, the Langevin equation 
reduces (reduced variables denoted by the same symbols) to

\begin{equation}
\frac{d^{2}x}{dt^{2}}=-\gamma(x)\frac{dx}{dt}
+cos x +F(t)
+\sqrt{\gamma(x) T}\xi(t),
\end{equation}
where $\gamma(x)=\gamma_0(1-\lambda sin(x+\phi))$ and with similar noise 
statistics for the reduced $\xi(t)$ as earlier.
Our purpose is to calculate $x(t)$ and $v(t)$ by solving the Langevin equation.
As mentioned earlier, so far no analytical solution could be found even 
for the average velocity of the particle when the external force $F(t)$ is 
time varying. However, when  $F(t)=F_0$, a constant, it is 
straightforward to extend the MCFM to 
obtain the drift velocity in the steady state situation. The method 
involves solving the Fokker-Planck equation\\
\begin{equation}
\frac{\partial W(x,v,t)}{\partial t} = {\cal{L}}_{FP}W(x,v,t)
\end{equation}
corresponding to the Langevin equation (2), where the FP operator
\begin{eqnarray}
\cal{L}_{FP} & = & -v\frac{\partial}{\partial x}+\gamma_0(1-sin(x+\phi))
\frac{\partial}
{\partial v}v - (cosx+F_0)\frac{\partial}{\partial v}+ \nonumber \\
      &    &\gamma_0T(1-\lambda sin (x+\phi))\frac{\partial^{2}}{\partial v^{2}}.
\end{eqnarray}
In the MCFM, the distribution $W(x,v,t)$ is expanded in terms of the function $\psi_n$,
\begin{equation}
W(x,v,t)=(2\pi T)^{\frac{-1}{4}}e{^\frac{-v^{2}}{4T}}\sum_{n=0}^
{\infty}C_n(x,t)\psi_n,
\end{equation}

where $\psi_n=\frac{(b^{\dagger})^n\psi_0}{\sqrt{n!}}$  are the eigen functions 
of the operator $\gamma(x)b^\dagger b$ and 
\begin{equation}
\psi_0=(2\pi)^\frac{-1}{4}T^\frac{-1}{4}e^\frac{-v^2}{4T}.
\end{equation}
$b=\sqrt{T}\frac{\partial}{\partial v}+\frac{1}{2}\frac{v}{\sqrt{T}}$,
$b^\dagger=-\sqrt{T}\frac{\partial}{\partial v}+\frac{1}{2}\frac{v}{\sqrt{T}}.$
In the steady state, the Fokker-Planck equation reduces to 
\begin{equation}
n\gamma(x)C_n(x)+\sqrt{n+1}DC_{n+1}(x)+\sqrt{n}\hat{D}C_{n-1}(x)=0,
\end{equation}
where $D=\sqrt{T}\frac{\partial}{\partial x}$ and 
$\hat{D}=\sqrt{T}\frac{\partial}{\partial x} + \frac{-cos x -F}{\sqrt{T}}$.
This is a series of recurring equations for $n=0,1,2,\ldots,N$. This series is 
truncated at $n=N$ which depends on the value of $\gamma$, $F$, etc and can be 
quite large. Since the potential as well as the friction coefficient are 
periodic, the mean particle velocity $<v>$ is given by 
$\int_{-\infty}^{\infty}dv\int_{0}^{2\pi}dx W(x,v)v$ which turns out to 
be a constant from the first, $n=0$, of the series of the equations (7). 
Solving these simultaneous equations to obtain $C_{n}(x)$ is not 
straightforward, however, and one may not obtain any stable solution for all 
values of $\gamma_0$, and $F$.

For the time dependent external force fields, one necessarily have to use 
numerical methods to solve the Langevin equation. For homogeneous inertial 
systems but with asymmetric periodic potentials, the ratchet effect has been
investigated recently by solving the Langevin equation numerically\cite{Mach}.
In the present work we have used stochastic form of the $4^{th}$ order 
Runge-Kutta method. Numerically, one can calculate position dispersion at 
various times and also the distribution of instantaneous velocities for the 
entire period of motion. The drift velocity can be obtained either by taking 
the mean position at a large time and dividing by the time: 
$<v>=\lim_{t \rightarrow \infty}\frac{<x(t)>}{t}$ or 
$<v> = \int_{-\infty}^{\infty}v\rho (v)dv$, where $\rho (v)$ is the velocity 
distribution. The $<\ldots>$ in 
$<v>=\lim_{t \rightarrow \infty}\frac{<x(t)>}{t}$ is the average evaluated 
over several realizations of the stochastic run whereas in the later case 
the (instaneous) velocity distribution is obtained for several realizations of 
the entire stochastic run. In the following we present results of our 
calculation first for the adiabatic case and then for the time-varying (finite 
frequency) field case.

\section{Numerical Results}
\subsection{Adiabatic Case}
\begin{figure}[]
\newpage
\epsfxsize=9cm
\centerline{\epsfbox{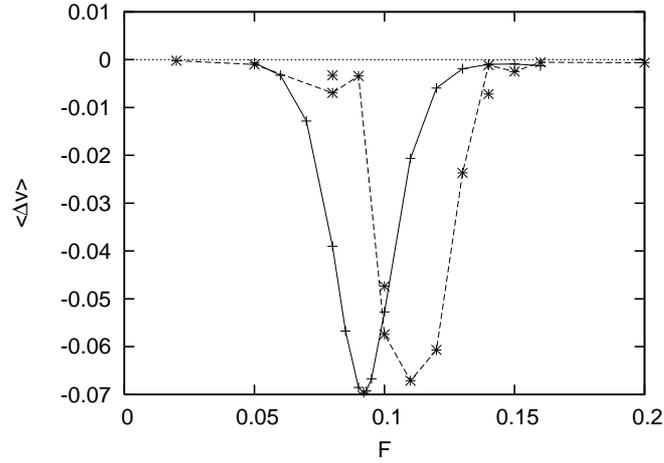}}
\caption{Shows the ratchet current $<\Delta v>$ as a function of $F$ using
MCFM ($\dagger$ joined by dashed lines) and the Langevin simulation method
($\ast$ joined by dashed lines) for $\gamma = 0.035$, $T = 0.4$, $\phi = 0.35$.
The dotted line for zero current is shown just to guide the eye.}
\label{fig.2}
\end{figure}

\begin{figure}[]
\epsfxsize=9cm
\centerline{\epsfbox{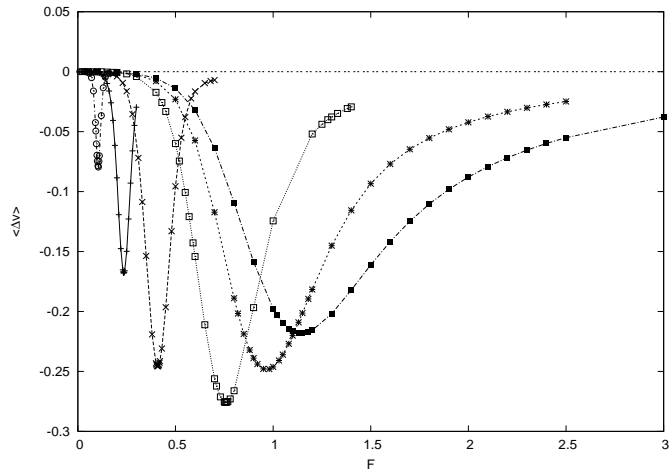}}
\caption{The variation of $<\Delta v>$ as a function of $F$ is shown for
$\gamma = $ 0.04 ($\circ$), 0.1 (+), 0.2 (x), 0.5
(open square), 0.75 ($\ast$) and 1.0 (solid squares)}
\label{fig.3}
\end{figure}

\begin{figure}[]
\epsfxsize=9cm
\centerline{\epsfbox{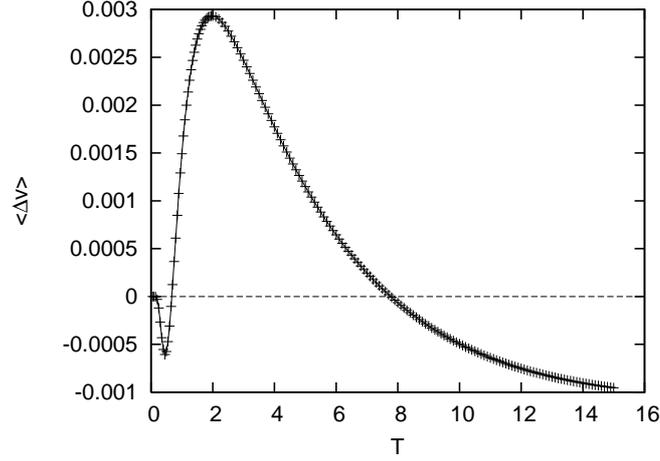}}
\caption{Shows variation of $<\Delta v>$ as a function of temperature $T$ for
$\gamma_0 = 0.1$ and $F_0=0.1$.}
\label{fig.4}
\end{figure}

\begin{figure}[]
\epsfxsize=9cm
\centerline{\epsfbox{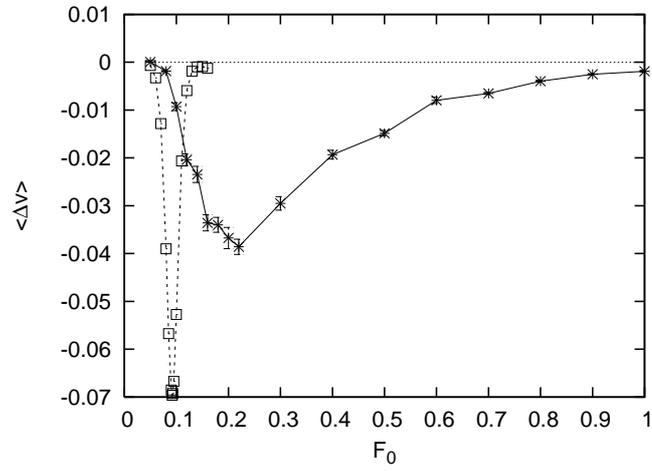}}
\caption{The ratchet current $<\Delta v>$ when the system was driven by a
square-wave field of amplitude $F_0$ are shown (+) with errorbars for
$\gamma_0 = 0.035$ and $T = 0.4$. The curve (with open box) is for the
adiabatic case for the same $\gamma_0$ and $T$}
\label{fig.5}
\end{figure}

\begin{figure}[]
\epsfxsize=9cm
\centerline{\epsfbox{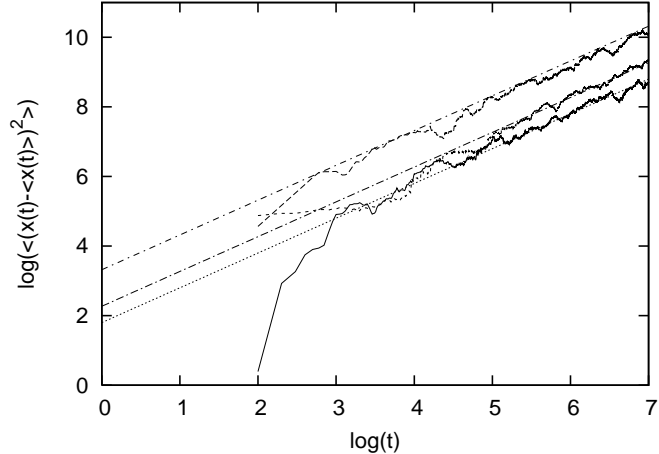}}
\caption{The position dispersions $<(\Delta x)^{2}>$ are plotted as a
function of time $t$ for $F_0 = 0.08$ (solid line), $0.3$ (dashed line)
and $0.6$ (small dashed line) for $\gamma = 0.035$ and $T = 0.4$. The straight
lines are drawn to indicate the average slope equal to 1 and intercepts
$1.8$, $3.2$ and $2.27$ respectively for the above $F_0$ values of the 
curves at large $t$ limit.}
\label{fig.6}
\end{figure}

\begin{figure}[]
\epsfxsize=9cm
\centerline{\epsfbox{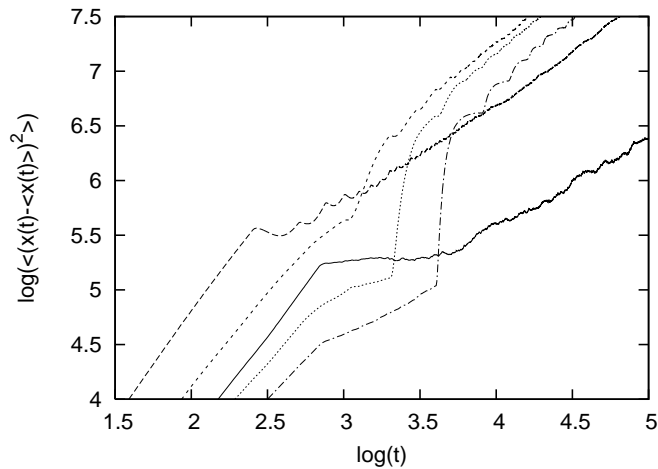}}
\caption{The position dispersions $<(\Delta x)^{2}>$ as a function of time
$t$ are shown for the square wave drive amplitude $F_0=0.2$ but with periods 
500 (long dashed line), 2000 (small dashed line), 4000 (dotted line), and 8000
(dashed-dotted line) for $\gamma = 0.035$, $T=0.4$, and $\phi = 0.35$. The
dispersions when a constant force $F_0=0.2$ is applied is shown by the solid
line. This curve is analogous to the homogeneous case of [7].}
\label{fig.7}
\end{figure}

The matrix continued fraction method is used to calculate the drift velocity,
(and hence mobilities), the average potential and kinetic energies, etc, as 
a function of $\gamma_0$,$F_0$ and T. However, in this paper we present only 
the drift velocities to illustrate the ratchet effect in inhomogeneous 
(non-uniform friction) inertial systems. As will be detailed in Ref \cite{Reen}
the expansion coefficients $C_n(x)$ are expressed in terms of their Fourier 
components $C_{n}^{q}$ which form column matrices  $\bf C_{n}$ for each n. By 
a clever transformation $\bf C_n$ and $\bf C_{n-1}$ are related. $\bf C_1$ 
being constant it can be found through the normalization condition,
$1=\int_{-\pi}^{\pi}C_0(x)dx$ and by evaluating the relational matrix (between 
$\bf C_{0}$ and $\bf C_{1}$) which depends on the Fourier componets of $V(x)$ 
and $\gamma (x)$ and $F$. A typical mean velocity evaluation may require $N$ 
as large as 1000 and the number of Fourier components typically as large as 
30, if at all a stable solution is to be found. The drift velocity, obtained
by taking the sum of mean velocities corresponding to the applied force 
$F=|F_0|$ and $F=-|F_0|$, as a function of $F$ is shown in Fig.2 for 
$\gamma_0=0.035$. This case of calculating the drift velocities or the ratchet 
currents, corresponding to the effect of application of a zero averaged drive 
field of frequency $\omega \rightarrow 0$, is referred to as the adiabatic case.
In the same figure we 
have also plotted drift velocities obtained by numerically solving the Langevin 
equation. Here we have presented the sum $\Delta \bar{v}$ of $\bar{v}$ for 
$F=\pm|F_0|$, 
where $\bar{v}=\lim_{t \rightarrow \infty}\frac{x(t)}{t}$ without the ensemble 
averaging. The ensemble averaging, however, does not qualitatively change the 
nature of the 
graph. The comparison of the two graphs shows that the Langevin simulation 
yields qualitatively similar result to what is obtained using MCFM. However, 
the peak ratchet currents occur at slightly different values of $F$, both being
at $F \ll 1$. Importantly, the ratchet current, in this adiabatic limit, 
results only in a very small range of $F$. In the following, for the 
adiabatic case, we present results obtained from MCFM calculations only.

From these calculations we find that the ratchet current do not change its 
qualitative behaviour for different $\gamma_0$. However, the narrow peaks occur 
at values of $F$ that depend on the values of $\gamma_0$ and $T$. As 
$\gamma_0$ is increased the peak position gradually shifts towards larger $F$
and also the peak width increases. The largest peak for $T=0.4$, occurs for 
$\gamma_0 = 0.4$. Also, at $T=0.4$ the ratchet current is in the negative 
direction (Fig.3). However, there is a distinct possibility of current 
reversals as a function of temperature $T$ as is seen in Fig.4. The figure also 
shows clearly that the net current could also peak as a function of the 
temperature $T$. It is worth reminding that the current $<\Delta v>$ here (in 
the adiabatic case) is just the sum of the average currents corresponding to 
$F = \pm |F|$. The average currents corresponding to $\pm |F|$ need not, 
however, be nonmonotonous individually. From Fig.3, there is a very clear 
indication that $<\Delta v(\gamma)>$ will also be non-monotonous.

Since there is no way we can calculate analytically the average currents for
time varying fields we present (numerical) Langevin simulation results 
for the non-adiabatic case.

\subsection{Non-adiabatic case}
In the finite (albeit small) frequency drive case, we consider a square wave 
force field 
where $F(t)=F_0$ for time interval $T_\Omega = 1000$ and in the next 
($T_\Omega = 1000$) interval\cite{JJ} the force is taken as $-F_0$ and 
repeated the 
procedure for a large number of periods (typically equal to 5,000 complete 
cycles) and 
the mean velocity is calculated. An average of this is taken over 
several realizations (typically 20), thus evaluating the ensemble averaged mean 
velocity or the net (ratchet) current. The obtained steady state current as a 
function of $F_0$ is plotted in Fig.5 for $\gamma_0 = 0.035$. For comparison 
the current obtained in the adiabatic case is plotted together. The range of
$F$ for which ratchet current can be obtained has broadened considerably in
comparison to the adiabatic case. The value of $F$ at which the peak current 
is obtained has also shifted to a higher value. In Fig.6, one can see that 
around $t \sim 10^{7}$, the position dispersion 
$<(\Delta x)^{2}>=<x^{2}(t)>-<x(t)>^{2} \sim t$ hinting at the approach to 
the diffusive steady state situation.

Fig.6 with  $<(\Delta x(t))^{2}> \sim t$ provides a method to calculate the 
diffusion constant $D$ in the large time limit. It is interesting to observe 
that in this limit $<(\Delta x)^{2}>$ shows a peaking behaviour as a function 
of F. For $\gamma = 0.035$, the peak ratchet current occurs close to where 
$<(\Delta x)^{2}>$ is maximum. It would, therefore, appear that ratchet 
current in this case is dispersive and not coherent. However, when the 
relative dispersion is plotted as a function of $F$, 
$\frac{<(\Delta x)^{2}>}{<x>^{2}}$ shows small values $<1$ in the range of $F$ 
where ratchet current is obtained. Thus, the obtained ratchet current is, 
indeed, a result of coherent motion of particles. Equivalently, the related 
Peclet number $\it{Pe}$ which is a measure of coherence of motion, turns out 
to be much larger than 2 indicating the particle motion giving rise to steady 
state ratchet current to be coherent. The motion at the intermediate 
(transient) time scales, however, shows a very different nature.

As mentioned earlier, Lindenberg, et. al.\cite{Linde} have shown that the 
motion of an ensemble of non-interacting particles in a periodic potential
together with a uniformly applied constant force showed dispersionless 
behaviour at intermediate time scales. They have set the time interval 
$[t_{min}<t< t_{max}]$ for a given applied force $F_0$ during which the 
system shows such an interesting behaviour. This behaviour is seen only for 
a certain range of $F_0$ values. A typical graph of $<(\Delta x(t))^{2}>$
on log-log scale is presented numerically in Fig.7. In this figure are also 
shown the effect of frequency of the dichotomously time varying (square wave) 
applied field of
same amplitude $F_0 = 0.2$. If the half period of the applied field over 
which the field remains constant before its sign is changed, 
$T_{\Omega} < t_{min}$, then after $T_\Omega$ the dispersion shows a dip
before again picking up. This is repeated in every $T_\Omega$. If, however,
$T_\Omega > t_{min}$ at $t = t_{min}$ the slope of $<\Delta x^{2}>$ changes
abruptly. At $t = T_\Omega$, $<(\Delta x)^{2}>$ begins to increase sharply 
before it appears to level itself up till it reaches $t = 2 T_\Omega$. The 
behaviour shown between $T_\Omega$ and $2 T_\Omega$ is seen to repeat every 
$T_\Omega$ for a long time. The frequency dependence of applied force on the 
dispersive behaviour of particle motion shows interesting possibilities of 
dispersion reduction as well as brief dispersionless motion.

\section{Discussion and Conclusion}
The present work gives numerical evidence of the possibility\cite{Mah1} of 
obtaining ratchet current in an inertial inhomogeneous symmetric periodic 
potential system subjected to a symmetrically driven periodic external field. 
The nonuniform temperature case has been investigated earlier by Blanter, et 
al.\cite{Blant}. We have calculated the net current from the average distance 
travelled after a long time. However, almost the same current is obtained from 
the (instantaneous) particle-velocity distribution over the entire sojourn 
time of the particles. But we find that the efficiency with which useful work 
can be derived from such a system is quite low and falls in the subpercentage 
range. This could be because the system inhomogeneity considered here is a 
feeble cause to generate net current without the asymmetric support of 
external forcings. Preliminary work, however, shows that the magnitude of 
ratchet current can be increased dramatically if, instead of time symmetric 
drive, time asymmetric drive with zero time averaged force per period is used.
Similar effect for overdamped case has been reported earlier\cite{Dan}.

The velocity distribution in this case shows an overwhelming asymmetry in 
favour of particle motion in the direction of current in contrast to the 
symmetric drive case where it shows almost symmetric velocity distribution. 
The width of the velocity distribution\cite{Mach} increases with the amplitude 
of the periodic driving. In conclusion, even though system inhomogeneity is a 
weak ingredient to obtain ratchet current it is a distinctive way of obtaining 
such a current.
\section{Acknowledgement}
MCM thanks A.~M.~Jayannavar for discussions. MCM and WLR acknowledge partial 
financial support from BRNS, DAE, Govt. of India. RR wishes to thank DST, 
Govt. of India for financial assistance (SR/FTP/PS-33/2004).


\begin{thebibliography}{99}
\bibitem{JJ}For an equivalent RCSJ model of Josephson junctions
the characteristic plasma frequency $\omega_J$ will be about $10^3$ times 
larger than the drive field frequency corresponding to $T_{\Omega}=1000$.
\end{thebibliography}
\end{document}